\documentstyle[prl,aps,epsfig,psfig,amsmath,amssymb,twocolumn,prbbib]{revtex}

\newcommand{\lapprox}{\lesssim}

\begin{document}
\draft
\twocolumn[\hsize\textwidth\columnwidth\hsize\csname  
@twocolumnfalse\endcsname

\title{Shrinking of a condensed fermionic cloud in a trap 
approaching the BEC limit}
\author{A. Perali, P. Pieri, and G.C. Strinati}
\address{Dipartimento di Fisica, UdR INFM,  Universit\`{a} di Camerino, 
I-62032 Camerino, Italy}

\date{\today}
\maketitle
\hspace*{-0.25ex}

\begin{abstract}
We determine the zero-temperature density profile of a cloud of fermionic
atoms in a trap subject to
a mutual attractive interaction, as the strength of the interaction is
progressively increased.
We find a significant decrease of the size of the atomic cloud as it
evolves from the weak-coupling (BCS)
regime of overlapping Cooper pairs to the strong-coupling (Bose-Einstein)
regime of non-overlapping
bound-fermion pairs.
Most significantly, we find a pronounced increase of the value of the density 
at the center of the trap (even  by an order of magnitude) when evolving 
between the two 
regimes. Our results are based on a generalized Thomas-Fermi approximation 
for the superfluid state, that covers continuously all coupling regimes.
\end{abstract}

\pacs{PACS numbers: 03.75.Hh, 03.75.Ss, 67.90.+z}
\hspace*{-0.25ex}
]
\narrowtext

The impressive experimental realization of Bose-Einstein condensation (BEC)
of ultracold bosonic atoms in a
trap over the last few years \cite{AK-02} has prompted the expectation that
also the properties of ultracold
fermionic atoms in a trap could be revealed in a regime of quantum
degeneracy \cite{DM-Jin-99}.
In addition, the possibility that the study of Fermi gases may lead to a
superfluid phase analogous to the
BCS state has added excitement to the field.
In particular, experimental proposals are under investigation for realizing
Cooper pairs with ultracold
dilute fermionic atoms in a trap, either via the enhanced interaction
between fermionic atoms with
a tunable scattering (Feshbach) resonance \cite{Jin-02} or via 
sympathetic cooling with
a mixture of fermions and bosons \cite{Inguscio-01}.

Given these rapid experimental advances, it seems timely to explore
theoretically the even more intriguing
question of the \emph{crossover\/} between the weak-coupling (BCS) regime
of overlapping Cooper pairs and the
strong-coupling (Bose-Einstein) regime of non-overlapping bound-fermion
pairs for a gas of ultracold Fermi
atoms in a trap, thus dealing simultaneously with two basic quantum
phenomena that share the same spontaneous
broken-symmetry (superfluid) behavior.
In particular, novel and interesting physics may result in the
intermediate-coupling regime where neither one
of the two limits is fully realized.

Detection of density profiles of ultracold trapped atoms is routinely used
to explore their degeneracy
properties.
This feature could also be explored to detect the continuous evolution
from Cooper pairs to
bound-fermion pairs as the strength of the attractive interaction between
fermions is increased, with
the expectation that the density profile shrinks upon approaching the
strong-coupling regime owing
to the diminished Fermi pressure.
In this respect, partial results have already been obtained as a
function of temperature in the broken-symmetry
phase for a single coupling toward the weak-coupling side of the BCS-BEC
crossover \cite{Chiofalo-02}, as well as
a function of coupling in the normal phase above the critical temperature
\cite{Griffin-02}.

In this paper, we calculate the density profile of a cloud of fermionic
atoms in a trap subject to a mutual
attractive interaction, as its strength is progressively
increased from weak to strong coupling.
To this end, we extend the coupled mean-field equations determining the gap
function and the chemical
potential for the BCS-BEC crossover problem of a homogeneous system to the
case of a \emph{spatially varying
gap function\/} in the presence of a harmonic trap, and study in a
systematic way the evolution of the density
profile at zero temperature where the mean-field equations are expected to
be reliable.
Our approach generalizes the fermionic Thomas-Fermi (TF) approximation to the
superfluid state at zero temperature for any coupling, and recovers the 
TF approximation for the composite bosons\cite{Pethick-Smith} in the 
strong-coupling limit.
We find a significant decrease of the size of the atomic cloud as the
strong-coupling limit is approached;
even more significantly, we find a pronounced increase of the value of 
the density at the center of the trap when evolving between the two regimes.
This increase can thus be regarded as a signature of the achieved
BCS-BEC crossover.
Numerical calculations for the density profile, gap function, and chemical
potential, as well as analytic
results for the radius of the cloud and related quantities are presented.

A central issue of the BCS-BEC crossover 
problem\cite{Ranninger} is to obtain for all 
coupling regimes {\em a single theory} expressed in terms of fermionic 
variables, 
which recovers known results for weakly-interacting fermions (on the 
weak-coupling side) and for weakly interacting bosons (on the strong-coupling 
side), thus providing an interpolation scheme over the intermediate coupling
regime. The treatment of the intermediate-coupling regime is necessarily an
approximate one, because in this regime a small parameter to control the 
approximations is lacking. For these reasons, the interpolation via a 
single fermionic theory represents at present the only manageable 
approach to the BCS-BEC crossover.
[Effects going beyond the BCS theory in weak coupling have been included
for the homogeneous case in Ref.\onlinecite{gorkov}
In the present treatment of the inhomogeneous case, we shall limit to consider
the crossover from BCS theory to the BEC limit, thus leaving out these
additional effects.]

The present calculations are based on a local-density approximation,
whereby the fermionic chemical potential
entering the BCS gap and density equations is replaced by a \emph{local
chemical potential\/} that adapts
to the spatial position in the harmonic trap.
We thus set $\mu({\mathbf r}) \, = \, \mu \, - \, V({\mathbf r})$, where $\mu$
is the equilibrium chemical
potential and $V({\mathbf r}) = m \omega^{2} \mathrm{r}^{2}/2$ is the
harmonic potential for the Fermi atoms at
position ${\mathbf r}$ in the trap ($m$ being the fermionic mass and
$\omega$ the characteristic trap frequency).
Correspondingly, the gap function takes the local value $\Delta({\mathbf r})$.
With the replacement of $\mu$ by $\mu({\mathbf r})$ and of $\Delta$ by
$\Delta({\mathbf r})$, the $s$-wave BCS
gap equation at zero temperaturen becomes ${\mathbf r}$
dependent and reads in dimensionless form:
\begin{equation}
-\frac{\pi}{k_{F}a_{F}} = \int_{0}^{\infty}\!d\tilde{E}\sqrt{\tilde{E}} 
\left( \frac{1}{\sqrt{(\tilde{E} - \tilde{\mu}(\tilde{r}))^{2} +
\tilde{\Delta}(\tilde{r})^{2}}}
- \frac{1}{\tilde{E}} \right).
\label{dimensionless-gap-equation}
\end{equation}
We have here introduced the dimensionless variables
$\tilde{r}=|{\mathbf r}|/R_{F}$,
$\tilde{\Delta}(\tilde{r})=\Delta(|{\mathbf r}|)/E_{F}$,
and $\tilde{\mu}(\tilde{r})=\mu(|{\mathbf r}|)/E_{F} = \tilde{\mu} -
\tilde{r}^{2}$
for a spherical trap, where $E_{F}=(3N)^{1/3} \omega$ ($\hbar = 1$
throughout) is the Fermi energy for
non-interacting fermions in the trap \cite{Pethick-Smith}, $R_{F}$ is
identified by the condition
$m \omega^{2} R_{F}^{2}/2=E_{F}$, and $\tilde{\mu}=\mu/E_F$.
We have further defined the Fermi wave vector $k_{F}$ by
$E_{F}=k_{F}^{2}/(2m)$.

The above equation has been obtained by considering a contact potential for
the attractive fermionic interaction
active between two equally populated fermionic species with different
internal (``spin'') states, and then
regularizing the ensuing ultraviolet divergence of the gap equation in
terms of the \emph{scattering length\/}
$a_{F}$ of the associated fermionic two-body problem
\cite{Randeria-93,Pi-S-98}.
This regularization, in fact, considerably simplifies dealing with the
BCS-BEC crossover and parametrizes the
interaction in terms of the measurable scattering length $a_{F}$.

Dealing with the BCS-BEC crossover requires us to supplement
the gap equation by the density equation
$N \, = \, \int \! d{\mathbf r} \, n({\mathbf r}) \, = \, \int \!
d\tilde{{\mathbf r}} \, \tilde{n}(\tilde{{\mathbf r}})$,
yielding
\begin{equation}
\frac{\pi}{24} =  \int_{0}^{\tilde{r}_{c}} \!\!d\tilde{r} \tilde{r}^{2}
                            \int_{0}^{\infty} \!\!d\tilde{E} \sqrt{\tilde{E}}
\left( 1 - \frac{\tilde{E} - \tilde{\mu}(\tilde{r})}
{\sqrt{(\tilde{E} - \tilde{\mu}(\tilde{r}))^{2} +
\tilde{\Delta}(\tilde{r})^{2}}} \right) .
\label{dimensionless-density-equation}
\end{equation}
Here, $\tilde{r}_{c}$ is the \emph{radius of the atomic cloud\/} defined by
the condition $\tilde{n}(\tilde{r})
=0$ for $\tilde{r}\ge\tilde{r}_{c}$ (which, in turn, requires
$\Delta(\tilde{r})=0$ for $\tilde{r}\ge\tilde{r}_{c}$
as a necessary condition).
It can be readily verified that $\tilde{r}_{c}$ is determined by the
equation $\tilde{\mu}(\tilde{r}_{c})=0$ for
$a_{F}<0$, yielding $\tilde{r}_{c}=\sqrt{\tilde{\mu}}$, and by the equation 
$\tilde{\mu}_{B}(\tilde{r}_{c})=0$ for
$a_{F}>0$, yielding
$\tilde{r}_{c}=\sqrt{\tilde{\mu} + (k_{F}a_{F})^{-2}}$.
In the above expression, we have introduced the local \emph{bosonic\/}
chemical potential 
$\mu_{B}(r) = \mu_{B} - V_{B}({\mathbf r})$, where $\mu_{B} = 2\mu \, + \,
\epsilon_{0}$ is the equilibrium 
bosonic chemical potential, $V_{B}({\mathbf r}) = 2V({\mathbf r}) = m_{B}
\omega^{2} \mathrm{r}^{2}/2$ is the harmonic
potential for the bosonic molecules of mass $m_{B}=2m$, and
$\epsilon_{0}=(m a_{F}^{2})^{-1}$ is the molecular binding
energy.
Contrary to Eq.~(\ref{dimensionless-gap-equation}),
Eq.~(\ref{dimensionless-density-equation}) couples different
positions in the trap, since obtaining a uniform chemical potential $\mu$
requires knowledge of
$\Delta({\mathbf r})$ and $\mu({\mathbf r})$ over the whole trap region.

The weak- and strong-coupling limits of 
Eqs.~(\ref{dimensionless-gap-equation}) 
and~(\ref{dimensionless-density-equation}) are essentially determined by the 
behavior of the chemical potential $\mu$, which about coincides with the Fermi
energy $E_F$ in the weak-coupling limit and with (minus half) the binding 
energy $\epsilon_0$ in the strong-coupling limit. 
These values represent, in fact, the energy required to extract one fermion 
from the system in the two limits, namely, from the Fermi sea and from a 
bound molecule, respectively.

In analogy to the homogeneous case \cite{mar-98}, \emph{analytic\/}
results can be obtained from the coupled equations
(\ref{dimensionless-gap-equation}) and
(\ref{dimensionless-density-equation}) both in the weak- and
strong-coupling
limits, defined, in the order, by the conditions $\tilde{\mu}(\tilde{r})>0$
and $\tilde{\Delta}(\tilde{r})\ll
\tilde{\mu}(\tilde{r})$, and $\tilde{\mu}(\tilde{r})<0$ and
$\tilde{\Delta}(\tilde{r})\ll|\tilde{\mu}(\tilde{r})|$
(for $0\le\tilde{r}\le\tilde{r}_{c}$).
In weak coupling, one obtains
$\tilde{\Delta}(\tilde{r}) =  (8 \tilde{\mu}(\tilde{r})/e^{2}) \,\, \exp
\left\{ - \pi /
(2 k_{F} |a_{F}| \sqrt{\tilde{\mu}(\tilde{r})}) \right\}$
with the typical essential singularity behavior as $k_{F}|a_{F}|
\rightarrow 0$, and the associated density profile
$\tilde{n}(\tilde{r}) = (8 N/ \pi^{2}) \left( 1 - \tilde{r}^{2} \right)^{3/2}$
which is the standard result within the TF approximation for
non-interacting fermions.
For the radius of the atomic cloud one gets correspondingly $r_{c}=R_{F}$.
In strong coupling, one obtains instead
$\tilde{\Delta}(\tilde{r})^{2} = 2 (\epsilon_{0}/E_{F})
\tilde{\mu}_{B}(\tilde{r})$
with the associated bosonic density
$n_{B}(r) = n(r)/2 = (m_{B}/4 \pi a_{B}) \mu_{B}(r)$
in terms of the original variables, where $a_{B}=2a_{F}$ identifies the 
bosonic scattering length.
The TF result for bosons \cite{Pethick-Smith} is thus
properly recovered by our
calculation in the strong-coupling limit, with a residual bosonic
interaction $U_{0} = 4 \pi a_{B} / m_{B}$
that matches previuos estimates for the homogeneous case.
[A more sophisticated theory for the homogeneous case~\cite{Pi-S-98}  
has shown that the correct value for the bosonic scattering length $a_B$ is 
smaller by about a factor of 3 than the value $2 a_F$.
Owing to the very slow dependence of the size of the Thomas-Fermi density
profile on $a_B$ in strong coupling (see below), this difference makes 
practically no change in this quantity.]
Using eventually the expression of $\mu_{B}$ for bosons in a trap
\cite{Pethick-Smith}, one gets
the value $\tilde{r}_{c} = \sqrt{\tilde{\mu}_{B}/2} = 0.69
(k_{F}a_{F})^{1/5}$  showing a nontrivial
dependence of $\tilde{r}_{c}$ on the coupling parameter in the
strong-coupling limit.

Recovering from Eqs.~(\ref{dimensionless-gap-equation}) and
(\ref{dimensionless-density-equation}) the TF results  both for non-interacting
fermions (in the weak-coupling-limit) and for weakly-interacting bosons (in the
strong-coupling limit) should not be surprising, since these equations can also
be obtained by generalizing the TF procedure for fermions to the superfluid 
state. To this end, one introduces an energy functional 
$E[n({\bf r}),\Delta({\bf r})]$ that depends on the local variables 
$n({\bf r})$ and $\Delta({\bf r})$, such that minimizing  
$E[n({\bf r}),\Delta({\bf r})]$ with
respect to variations of $\Delta({\bf r})$ and $n({\bf r})$ (subject to the 
constraint $N=\int d{\bf r} \, n({\bf r})$) leads to 
Eqs.~(\ref{dimensionless-gap-equation}) and 
(\ref{dimensionless-density-equation}), respectively. Within the present 
approach, this functional is given by the expression
\begin{eqnarray}
& &E[n({\bf r}),\Delta({\bf r})] = \int\!d{\bf r}\left\{
n({\bf r})V({\bf r}) - \frac{m}{4\pi a_F}\Delta({\bf r})^2  
\right.\nonumber\\
& &\left.+\int\!\frac{d{\bf k}}{(2 \pi)^3}\left[\epsilon_{\bf k} -
\frac{\epsilon_{\bf k}(\epsilon_{\bf k}-\lambda)+
\Delta({\bf r})^2}{\sqrt{(\epsilon_{\bf k}-\lambda)^2+\Delta({\bf r})^2}}+
\frac{\Delta({\bf r})^2}{2 \epsilon_{\bf k}}\right] 
\right\}
\label{functional}
\end{eqnarray}
where $\epsilon_{\bf k}=k^2/(2m)$ and $\lambda$ is here an implicit function of
$n({\bf r})$ and $\Delta({\bf r})$ via the equation
\begin{equation}
n({\bf r})=\int\!\frac{d{\bf k}}{(2 \pi)^3}\left[1 -\frac{\epsilon_{\bf k}
-\lambda}{\sqrt{(\epsilon_{\bf k}-\lambda)^2+\Delta({\bf r})^2}}
\right]
\;.
\end{equation} 
In particular, in the strong-coupling limit (whereby $n({\bf r})\propto 
\Delta({\bf r})^2$) the functional~(\ref{functional}) reduces to
\begin{equation}
E[n({\bf r})]=\int\!d{\bf r}\left[\frac{U_0}{8}n({\bf r})^2+\left(V({\bf r})-
\frac{\epsilon_0}{2}\right)n({\bf r})\right]
\label{functstr}
\end{equation} 
from which the bosonic TF expression for $n_{{\rm B}}({\bf r})$ can be 
obtained by minimizing with respect to $n({\bf r})$ and taking the constraint
into account.
[The absence of the kinetic energy term in the bosonic TF approximation results
in the present approach by the neglecting of the center-of-mass motion of the
pairs in the expression~(\ref{functional}) for the energy.] 
This results implements at the level of the TF approximation the density 
functional theory for superconductors described in Ref.~\onlinecite{kohn}.

The rescaled equations (\ref{dimensionless-gap-equation}) and
(\ref{dimensionless-density-equation}) depend
\emph{only\/} on the dimensionless coupling parameter $(k_{F}a_{F})^{-1}$
and not on the total number of particles $N$.
This is due to the fact that, as expected, the kinetic energy is not
properly taken into account in the
present local-density treatment that reduces to the TF approximation for
fermions in weak coupling and for bosons
in strong coupling.
What is missing in our equations is the presence of the length scale of the
harmonic trap,
which for a bosonic molecule equals $a_{\mathrm{osc}}^{B} = (m_{B}
\omega)^{-1/2}$ and to which there would
be associated the $N$-dependent dimensionless ratio
$R_{F}/a_{\mathrm{osc}}^{B} = 2.402 \, N^{1/6}$.
The physical lower bound $r_{c} = a_{\mathrm{osc}}^{B}$, however, can never be
reached in practice, since it would
correspond to the condition $(k_{F}a_{F})^{-1} = 12.5 \, N^{5/6}$, yielding
the totally unrealistic value
$a_{F} \sim 10^{-3} a_{0}$ ($a_{0}$ being the Bohr radius) in the case of a
fermionic atom like $^{40}K$
for $N \sim 10^{6}$ and a typical trap value $\omega \sim 10^{3} s^{-1}$.

By a similar token, it can be shown that the local-density approximation for
the trapped Fermi gas holds provided $k_F R_F\gg 1$ (such that the energy 
quantization in the trap is irrelevant with respect to $E_F$) and 
$\Delta({\bf r}=0)/\omega\gg 1$ (such that the pairs are well contained within
the trap)\cite{Bruun}. When $N\sim 10^6$, $k_F R_F \sim 10^2$ while 
$\Delta({\bf r}=0)/\omega\sim 10$ already for the weak-coupling value 
$(k_F a_F)^{-1}=-2$.

Figure~1 shows the density profile $\tilde{n}(\tilde{r})$ along the radial 
coordinate (in units of
$R_{F}$) for several values
of the coupling parameter $(k_{F}a_{F})^{-1}$.
The TF profiles for non-interacting fermions as well as for interacting
point-like bosons (with $(k_{F}a_{F})^{-1}=5$)
are also shown for comparison.
Note from the figure that the non-interacting fermion density profile is
recovered in weak coupling and the
interacting point-like boson density profile is recovered in strong
coupling, respectively.
Note also the shrinking of the size of the atomic cloud as it evolves from
the weak-coupling to the strong-coupling regime, the critical radius 
$\tilde{r}_{c}$ being reduced by
about a factor of $2$ across the
crossover region.
Correspondingly, the density at the center of the trap increases
significantly (by about a factor of $6$ for
the coupling values shown in the figure).

\begin{figure}
\centerline{\psfig{figure=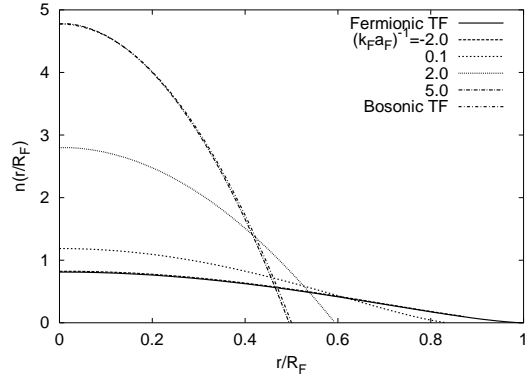,width=5.1cm,angle=-90}}
\vspace{0.1cm}
\caption{Density profile (divided by $N$) vs the radial
coordinate for several values of
the coupling parameter $(k_{F}a_{F})^{-1}$.}
\label{figure1}
\end{figure}

Figure~2 shows the radius $\tilde{r}_{c}$ of the atomic cloud (full line)
vs the coupling
parameter, as determined by entering the numerical values of the chemical
potential (obtained from the
solution of Eqs.~(\ref{dimensionless-gap-equation}) and
(\ref{dimensionless-density-equation}))
into the analytic expressions of $\tilde{r}_{c}$ reported previously.
Three distinct regions of the coupling parameter are identified from this plot:
(i) A weak-coupling region for $(k_{F}a_{F})^{-1} \lapprox -1$ where
$\tilde{r}_{c}$ is almost coupling-independent
and equal to $1$;
(ii) An intermediate-coupling region for $-1 \lapprox (k_{F}a_{F})^{-1}
\lapprox 1$ where $\tilde{r}_{c}$
rapidly decreases for increasing coupling, being reduced by about $35\%$
when $(k_{F}a_{F})^{-1}=1$;
(iii) A strong-coupling region for $1 \lapprox (k_{F}a_{F})^{-1}$ where
$\tilde{r}_{c}$ coincides with
the analytic expression of the bosonic TF theory (dashed line).
For completeness, we also show in the inset the value of the density
$\tilde{n}(\tilde{r}=0)$ at the center of the trap vs the
coupling parameter (full line).
Following a flat behavior in the weak-coupling regime, a quite rapid
increase of $\tilde{n}(\tilde{r}=0)$ occurs in the
intermediate-coupling regime, approaching eventually the power-law
dependence $\tilde{n}(\tilde{r}=0) \propto (k_{F}a_{F})^{-3/5}$
obtained analytically in the strong-coupling regime (dashed line).

The chemical potential is finally shown in Fig.~3 (full line) vs the
coupling parameter.
The behavior of the chemical potential obtained from the expression $\mu
= (\mu_{B} - \epsilon_{0})/2$ in
strong coupling (with the value of $\mu_B$ for bosons in a 
trap~\cite{Pethick-Smith}) is also shown for comparison (dashed line).
\begin{figure}
\centerline{\psfig{figure=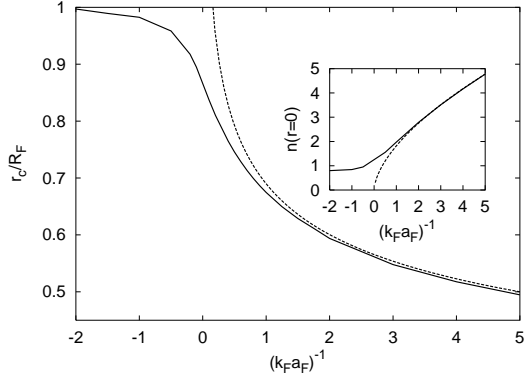,width=5.1cm,angle=-90}}
\vspace{0.1cm}
\caption{Radius of the atomic cloud (full line) and its
strong-coupling approximation (dashed line)
vs the coupling parameter $(k_{F}a_{F})^{-1}$.
Inset: corresponding values of the dimensionless density at the center
of the trap.}
\label{figure2}
\end{figure}
\noindent
In the inset, the gap parameter $\tilde{\Delta}(\tilde{r}=0)$ 
at the center of the trap (full line) is
shown vs the coupling parameter and compared with its strong-coupling
approximation (dashed line)
(the value for the three-dimensional homogeneous case \cite{mar-98}
(dashed-dotted line) is also shown).
Note the marked difference between the trapped and homogeneous cases, with
the value for the trapped case
increasing much faster than for the homogeneous case as the coupling increases.
Specifically, in strong coupling the analytical arguments discussed above yield
$\tilde{\Delta}(\tilde{r}) =(2\sqrt{2}/k_{F}a_{F})
\sqrt{0.476(k_{F}a_{F})^{2/5} - \tilde{r}^{2}}$,
producing the value $\tilde{\Delta}(0) = 1.951 (k_{F}a_{F})^{-4/5}$ at the
center of the trap;
for the homogeneous case, on the other hand, $\Delta/E_{F} = 1.303
(k_{F}a_{F})^{-1/2}$.

As the data of Figs.~1-3 for the various physical quantities $n(r)$,
$\Delta(r)$, $r_{c}$, and $\mu$ 
show, the BCS-BEC crossover takes place in practice over the rather narrow 
range $-1 \lapprox (k_{F}a_{F})^{-1} \lapprox 1$ of the coupling parameter.
This remark suggests that it would be
especially interesting to explore experimentally this
intermediate-coupling region, where deviations from the purely fermionic and 
bosonic behaviors occur.

In conclusion, we have shown that the density profile for superfluid 
fermionic atoms in a trap, exhibits an interesting
shrinking of the size of the atomic cloud as the strength of the attraction
increases.
This behavior should permit one to decide whether the superfluid behavior
is either of the BCS type with strongly
overlapping Cooper pairs or of the BEC type with non-overlapping
bound-fermion pairs.
\begin{figure}
\centerline{\psfig{figure=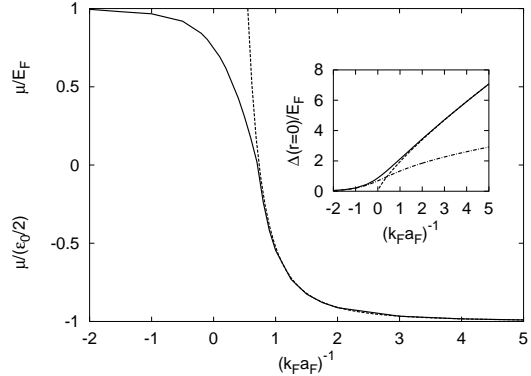,width=5.1cm,angle=-90}}
\vspace{0.1cm}
\caption{Chemical potential in units of $E_{F}$ when $\mu > 0$
and of $\epsilon_{0}/2$ when $\mu < 0$
(full line) and its strong-coupling approximation
(dashed line) vs the coupling parameter $(k_{F}a_{F})^{-1}$.
Inset: gap parameter at the center of the trap (full
line) and its strong-coupling approximation
(dashed line) vs the coupling parameter
(the value for 3D homogeneous case (dashed-dotted line)
is also shown for comparison).}
\label{figure3}
\end{figure}
\noindent

We are indebted to M. Inguscio and G. Modugno for discussions.



\end{document}